\documentclass[sigconf,natbib=true]{acmart}
\AtBeginDocument{%
  \providecommand\BibTeX{{%
    \normalfont B\kern-0.5em{\scshape i\kern-0.25em b}\kern-0.8em\TeX}}}

\usepackage[utf8]{inputenc}
\DeclareUnicodeCharacter{1EAB}{\^a}
\DeclareUnicodeCharacter{1EBF}{\^e}

\copyrightyear{2025}
\acmYear{2025}
\setcopyright{cc}
\setcctype{by}
\acmConference[SIGIR '25]{Proceedings of the 48th International ACM SIGIR
Conference on Research and Development in Information Retrieval}{July 13--18,
2025}{Padua, Italy}
\acmBooktitle{Proceedings of the 48th International ACM SIGIR Conference on Research and Development in Information Retrieval (SIGIR '25), July 13--18,
2025, Padua, Italy}
\acmDOI{10.1145/3726302.3730353}
\acmISBN{979-8-4007-1592-1/2025/07}

\usepackage[utf8]{inputenc} 
\usepackage{amsfonts}       
\usepackage{graphicx}
\usepackage{amsthm}
\usepackage{subfigure}
\usepackage{multirow}
\usepackage{enumitem}
\usepackage{bbding}
\usepackage{colortbl}
\usepackage{tcolorbox}
\usepackage{textcomp} 

\begin{document}

\title{NExT-Search: Rebuilding User Feedback Ecosystem for \\ Generative AI Search}

\author{Sunhao Dai}
\affiliation{
  \institution{\mbox{Gaoling School of Artificial Intelligence}\\Renmin University of China}
    \city{Beijing}
  \country{China}
  }
\email{sunhaodai@ruc.edu.cn}

\author{Wenjie Wang}
\authornote{Corresponding authors}
\affiliation{
  \institution{University of Science and Technology of China}
    \city{Hefei}
  \country{China}
  }
\email{wenjiewang96@gmail.com}

\author{Liang Pang}
\authornotemark[1]
\affiliation{
  \institution{CAS Key Laboratory of AI Safety\\Institute of Computing Technology Chinese Academy of Sciences}
    \city{Beijing}
  \country{China}
  }
\email{pangliang@ict.ac.cn}

\author{Jun Xu}
\affiliation{%
  \institution{\mbox{Gaoling School of Artificial Intelligence}\\Renmin University of China}
 \city{Beijing}
  \country{China}
}
\email{junxu@ruc.edu.cn}

\author{See-Kiong Ng}
\affiliation{
  \institution{\mbox{National University of Singapore}}
  \country{Singapore}
  }
\email{seekiong@nus.edu.sg}

\author{Ji-Rong Wen}
\affiliation{
  \institution{\mbox{Gaoling School of Artificial Intelligence}\\Renmin University of China}
    \city{Beijing}
  \country{China}
  }
\email{jrwen@ruc.edu.cn}

\author{Tat-Seng Chua}
\affiliation{
  \institution{\mbox{National University of Singapore}}
  \country{Singapore}
  }
\email{dcscts@nus.edu.sg}

\renewcommand{\authors}{Sunhao Dai, Wenjie Wang, Liang Pang, Jun Xu, See-Kiong Ng, Ji-Rong Wen and Tat-Seng Chua}
\renewcommand{\shortauthors}{Sunhao Dai, Wenjie Wang, Liang Pang, Jun Xu, See-Kiong Ng, Ji-Rong Wen and Tat-Seng Chua}
\renewcommand{\shorttitle}{NExT-Search: Rebuilding User Feedback Ecosystem for Generative AI Search}

\begin{abstract}
Generative AI search driven by large language models (LLMs) is reshaping information retrieval by offering end-to-end answers to complex queries, reducing users’ reliance on manually browsing and summarizing multiple web pages. However, while this paradigm enhances convenience, it disrupts the feedback-driven improvement loop that has historically powered the evolution of traditional Web search. Web search can continuously improve their ranking models by collecting large-scale, fine-grained user feedback (e.g., clicks, dwell time) at the document level. In contrast, generative AI search operates through a much longer search pipeline—spanning query decomposition, document retrieval, and answer generation—yet typically receives only coarse-grained feedback on the final answer. This introduces a \emph{feedback loop disconnect}, where user feedback for the final output cannot be effectively mapped back to specific system components, making it difficult to improve each intermediate stage and sustain the feedback loop.

To address this limitation, we envision \textbf{NExT-Search}, a next-generation paradigm designed to reintroduce fine-grained, process-level feedback into generative AI search. NExT-Search integrates two complementary modes: \textbf{User Debug Mode}, which allows engaged users to intervene at key stages—such as refining query decomposition, rating retrieved documents, and editing initial generated responses—and \textbf{Shadow User Mode}, where a personalized user agent simulates user preferences and provides AI-assisted feedback for less interactive users. As these feedback signals serve as valuable resources for refining the whole search pipeline, we also introduce a feedback store mechanism that encourages users to share and monetize their debugging efforts, further incentivizing participation. Furthermore, we envision how these feedback signals can be leveraged through \textbf{online adaptation}, which refines current search outputs in real-time, and \textbf{offline update}, which aggregates interaction logs to periodically fine-tune query decomposition, retrieval, and generation models. By restoring human control over key stages of the generative AI search pipeline, we believe NExT-Search offers a promising direction for building feedback-rich AI search systems that can evolve continuously alongside human feedback.
\end{abstract}

\ccsdesc[500]{Information systems~Users and interactive retrieval}

\keywords{Generative AI Search, User Feedback, Large Language Model}

\maketitle

\section{Introduction}

\begin{figure*}[t]  
    \centering    
    \includegraphics[width=1\linewidth]{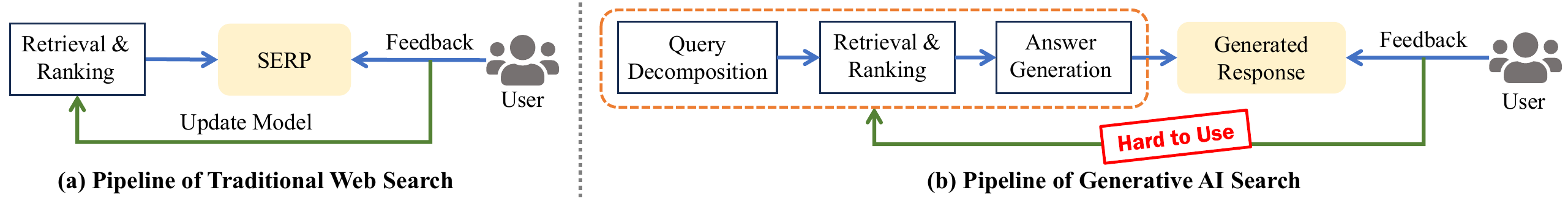}
    \caption{Comparison of the paradigm of traditional web search engines and generative AI search engines. (a) Web search retrieves and ranks results, presenting them as a Search Engine Results Page (SERP), where user feedback on the document level can be directly leveraged to update the ranking model. (b) Generative AI search automates multiple steps to generate direct answers, but its extended pipeline complicates the effective use of user feedback for refining individual components.}
    \label{fig:pipeline_compare}  
\end{figure*}

Search engines have served as a primary gateway to information access for decades, helping users address an enormous spectrum of information needs~\cite{manning2009introduction, chowdhury2010introduction, croft2010search}.
Despite continuous advances in ranking algorithms~\cite{burges2005learning, xu2018deep}, user interface design~\cite{hearst2009search}, and large-scale log data analysis~\cite{joachims2002optimizing, craswell2008experimental}, recent estimates suggest that nearly half of all Web search queries fail to yield relevant results~\footnote{\url{https://blogs.microsoft.com/blog/2023/02/07/reinventing-search-with-a-new-ai-powered-microsoft-bing-and-edge-your-copilot-for-the-web/}}. Many of these queries lie on the \emph{complex} end of the search spectrum: they require users to break down their search goal and iteratively check different returned search results, aggregating disparate pieces of information in a manual, time-consuming process~\cite{white2024advancing, hassan2014supporting}.

Recently, large language model (LLM)--driven generative AI search systems have promised to address these more complex queries in an end-to-end fashion~\cite{zhu2023large, li2024survey, xiong2024search, zhai2024large}. Users can now pose open-ended or creative requests (e.g., \emph{``Plan a trip to attend SIGIR 2025''}), and the generative AI search automatically parses them, retrieves relevant documents in smaller chunks, and then aggregates and synthesizes the extracted information into a cohesive answer~\cite{gao2023retrieval, fan2024survey}. Thus, by automating many of the steps traditionally performed by users in Web search, generative AI search \emph{expands the task boundary} of search, enabling users to solve complex, multi-faceted queries with reduced effort and cognitive load.

However, by first revisiting and comparing the foundational paradigms of both traditional Web search and advanced generative AI search in Section~\ref{sec:revisiting} and Figure~\ref{fig:pipeline_compare}, we find that while the paradigm of generative AI search introduces notable advantages, it also loses a critical component that has historically driven the success of traditional search: \textbf{the user feedback ecosystem}.  
\begin{itemize} [leftmargin=*]
    \item \textbf{Traditional Web search} thrives on a \emph{feedback-driven improvement loop}, where ranking models continuously evolve by leveraging large-scale user feedback on search results. Users can provide feedback at the document level, such as clicks, dwell time, and bounce rates. These fine-grained signals serve as direct supervision for refining retrieval and ranking models, enabling search engines to iteratively enhance result relevance and improve search quality over time~\cite{kelly2003implicit, agichtein2006improving, joachims2002optimizing, joachims2017accurately}.  

    \item \textbf{Generative AI search} operates through a much longer pipeline that directly synthesizes answers, significantly reducing human control over the search process. Users can only provide feedback in coarse-grained forms to the final generated response, such as simple likes/dislikes or written comments~\cite{white2024advancing}. This feedback loop disconnect prevents effective attribution of user dissatisfaction to specific pipeline components—whether query decomposition, retrieval, or answer generation—making it difficult to improve intermediate stages. 
\end{itemize}
Thus, while generative AI search systems such as Microsoft Bing Copilot~\footnote{\url{https://copilot.microsoft.com}} and Perplexity AI~\footnote{\url{https://www.perplexity.ai}} gained remarkable traction in their early launches, they still account for only a small share of the global search engine market~\footnote{\url{https://gs.statcounter.com/search-engine-market-share}}. Addressing these limitations is crucial for enabling generative AI search engines to achieve scalable, iterative optimization while retaining the benefits of automation.

To this end, we envision a new paradigm called \textbf{NExT-Search}, aimed at restore—and potentially enhance—the feedback-driven ecosystem for generative AI search.
The central idea is to incorporate two complementary modes of interaction: an active \textbf{User Debug Mode}, which enables users to engage with and refine each stage of the search pipeline, and a passive \textbf{Shadow User Mode}, which leverages a personalized user agent to simulate feedback when users prefer minimal involvement.
In the User Debug Mode, users can examine and modify different stages of the generative search workflow—such as query decomposition, document retrieval, and answer generation—providing fine-grained feedback as needed. 
In contrast, the Shadow User Mode engages a personalized user agent that learns from past interactions and user profiles to provide AI-assisted feedback throughout the pipeline, thereby reducing user effort while maintaining the flow of valuable supervision signals.
Together, these two modes offer a potential path toward gathering higher-quality feedback at various junctures in the pipeline.

Building on the fine-grained feedback envisioned in the NExT-Search paradigm, we further discuss how such feedback could be leveraged through two complementary update strategies. First, \textbf{online adaptation} enables generative AI search to refine current sessions in real time—for instance, re-ranking documents or partially regenerating answers in response to new feedback from either users or a personalized agent. Second, \textbf{offline update} aggregate logs from multiple interactions to periodically retrain or fine-tune crucial pipeline modules, preserving the iterative feedback-driven improvement loop that once propelled the success of traditional Web search. To further incentivize user participation, we envision a \textbf{Feedback Store} mechanism that enables users to share and potentially monetize their debugging contributions, making feedback not only a technical asset but also a user-valued commodity.
In this way, our NExT-Search paradigm has the potential to not only enhance search quality and user experience but also foster a mutually beneficial ecosystem where users are rewarded for their engagement while driving continuous model improvements.

Finally, as a perspective paper, our aim is to spark new thinking about why and how user feedback should be reimagined in the evolving paradigm of generative AI search. To this end, we also outline three promising research directions: 
building personalized user simulators to generate realistic AI-assisted feedback at scale, designing human-centric interfaces to make pipeline-level debugging accessible and efficient, and developing learning algorithms that can effectively leverage both human and AI-assisted feedback. Together, these directions chart a roadmap for building next-generation AI search systems that can continuously evolve in tandem with human feedback.

In summary, the main contributions of this paper are as follows:

$\bullet$ We provide a systematic analysis of the transition from traditional Web search to generative AI search, highlighting the core reasons why generative AI search has struggled to achieve large-scale success—chiefly, the loss of rich user feedback loops.

$\bullet$  We envision a new paradigm for generative AI search, called NExT-Search, which aims to rebuild the user feedback ecosystem with two modes: a \emph{User Debug Mode} that enables step-by-step user debugging across the pipeline and a \emph{Shadow User Mode} that simulates user feedback using personalized user agents to support minimal-interaction users.

$\bullet$ We outline how fine-grained feedback can be utilized through both online adaptation and offline model updates, and propose a feedback store to incentivize user participation, laying the foundation for a sustainable and self-improving search ecosystem.

\section{Web Search vs. Generative AI Search} \label{sec:revisiting}

In this section, we first revisit the pipelines of both traditional Web search and advanced generative AI search. By comparing the two pipelines, we underscore how the shift from a retrieval system with web page ranking lists to an end-to-end natural language answer-generating system leads to both gains and losses.

\subsection{The Pipeline of Traditional Web Search} \label{sec:websearch_pipeline}

Traditional Web search engines, such as Google and Bing, have evolved over decades of research and industrial practice in information retrieval (IR)~\cite{manning2009introduction, chowdhury2010introduction, croft2010search}. As shown in Figure\ref{fig:pipeline_compare}(a), the core pipeline typically consists of the following stages:

\textbf{(1) Retrieval \& Ranking.}
Upon receiving a user query (typically expressed as a set of keywords), the search engine initiates the retrieval process by identifying candidate documents from its index. This process leverages a hybrid approach that combines traditional keyword matching techniques (e.g., BM25~\cite{robertson2009probabilistic}) with advanced semantic or embedding-based matching methods~\cite{li2014semantic, guo2016deep, xu2018deep} (e.g., DSSM~\cite{huang2013learning}). Following retrieval, the system usually applies a learning-to-rank (LTR) model~\cite{liu2009learning} that integrates multiple features, including term-matching signals, link-based authority metrics (e.g., PageRank), user behavior patterns (e.g., click-through rates), temporal relevance indicators, and other contextual signals.

\textbf{(2) Result Delivery \& User Feedback.}
The retrieval and ranking stage then produces a Search Engine Results Page (SERP)~\cite{hochstotter2009users}, which generally lists tens or hundreds of hyperlinks accompanied by titles and snippets~\cite{kelly2015many}. Users then interact with and provide implicit feedback on these results, primarily through measurable engagement metrics such as click-through patterns, dwell time duration, bounce rates, and other interaction signals, all of which are systematically captured in real-time logs and subjected to continuous computational analysis~\cite{kelly2003implicit, agichtein2006improving, joachims2002optimizing, joachims2017accurately}.

\textbf{(3) Model Update.}
The large-scale feedback signals collected from user interactions then serve as direct supervision for refining ranking and retrieval models. User engagement at the document level—such as clicks on relevant results and skips on irrelevant ones—acts as a learning signal that continuously informs the ranking function, optimizing document relevance estimation.

The above stages drive an iterative refinement process for traditional web search known as the \textbf{data flywheel}~\cite{levene2011introduction, halavais2017search}: the more interactions occur, the more feedback is accumulated; the more training data is available, the better the ranking model becomes; and as search quality improves, more users engage with the system, further reinforcing the cycle. This feedback-driven improvement loop has been fundamental to the scalability and success of traditional Web search engines (e.g., Google and Bing).

\subsection{The Pipeline of Generative AI Search}
While traditional Web search has been highly effective in retrieving relevant documents, it struggles with complex or multi-step queries. Users often need to iteratively refine their searches, aggregate information from multiple sources, and manually synthesize answers to fulfill their information needs. 
Recent advancements in LLMs have given rise to the emergence of generative AI search engines, enabling them to go beyond hyperlink retrieval and provide end-to-end, synthesized answers tailored to user queries~\cite{zhu2023large, li2024survey, xiong2024search}. Instead of relying on users to extract and combine information, generative AI search automates the entire process—from decomposition and retrieval to synthesis—providing more direct responses. 
A representative pipeline of this process is illustrated in Figure~\ref{fig:pipeline_compare}(b).

 \textbf{(1) Query Decomposition.}
Generative AI search systems often adopt a retrieval-augmented generation (RAG) framework~\cite{liu2023evaluating, gao2023retrieval, white2024advancing}, which begins with decomposing complex user queries into one or more coherent sub-queries. This decomposition enables the system to iteratively refine and address partial information needs, ensuring a more precise and contextually relevant final response. By breaking down multifaceted queries, the system can better align with user intent and improve the overall search experience.

\textbf{(2) Retrieval \& Ranking.}
Following task decomposition, the system retrieves and ranks relevant passages for each sub-query, akin to traditional search engines. However, generative AI search typically operates at a finer granularity, retrieving and ranking text chunks or paragraphs rather than entire documents~\cite{gao2023retrieval}. This approach ensures that the retrieved evidence is both semantically aligned with the user's intent and sufficiently detailed to support high-quality answer generation.

 \textbf{(3) Answer Generation.}
The core of generative AI search lies in its ability to synthesize retrieved evidence into a coherent, natural-language response using LLMs~\cite{li2024survey}. The LLM integrates information from multiple sources, producing fluent and contextually rich answers that directly address the user's query. To enhance transparency and trustworthiness, the system may optionally include citations or references to the original sources, allowing users to verify the information and trace its provenance~\cite{liu2023evaluating}.

 \textbf{(4) Result Delivery \& User Feedback. }
Generative AI search engines present their outputs through conversational or chat-style interfaces, offering users a summarized and digestible answer in a single interaction. 
However, the user feedback in generative AI search is often more sparse than in traditional systems, typically limited to simple likes/dislikes or brief written comments to the final answer, posing challenges for fine-grained system improvement. 

Unlike traditional Web search, which benefits from fine-grained feedback at the document level, the coarse-grained feedback from generative AI search primarily presents a fundamental challenge: dissatisfaction with the final answer does not directly indicate whether errors originated from query decomposition, retrieval, or answer generation, making it difficult to pinpoint and correct specific weaknesses in the pipeline.

\begin{figure*}[t]  
    \centering    
    \includegraphics[width=0.95 \linewidth]{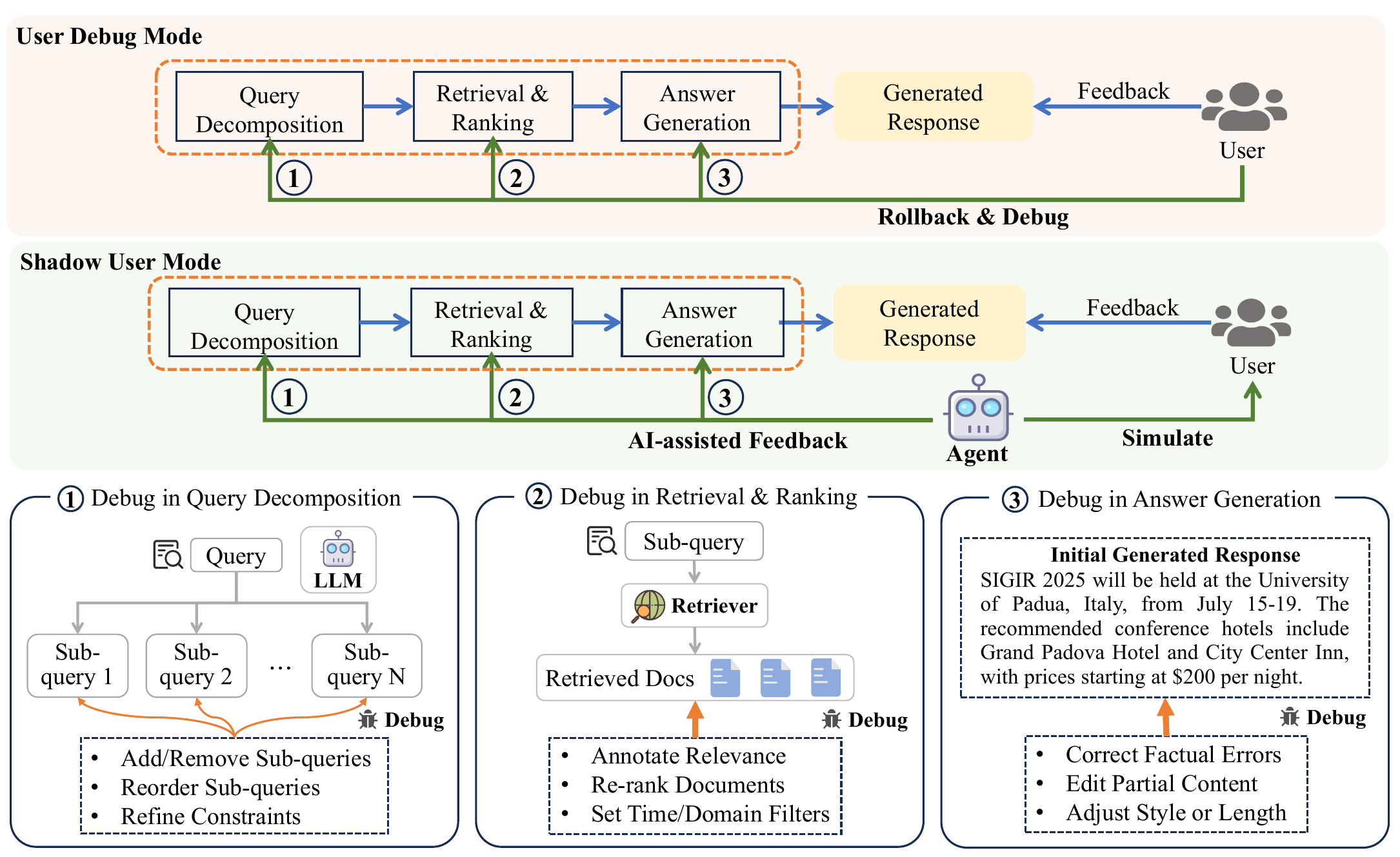}
    \caption{Illustration of our proposed NExT-Search paradigm. NExT-Search introduces a dual feedback mechanism to enhance generative AI search: In \emph{User Debug Mode}, users can intervene at different stages—query decomposition, retrieval, and answer generation—to refine search results with granular feedback. In \emph{Shadow User Mode}, a personalized agent simulates user behavior to assist in providing feedback with minimal user interaction, reducing user engagement costs.}
    \label{fig:NExT-Search_pipeline}  
\end{figure*}

\subsection{Comparison and Analysis} \label{sec:pipeline_comparison_analysis}

After illustrating the pipelines of both traditional Web search and generative AI search, we now summarize the key shifts in these two paradigms, highlighting the gains in usability and efficiency, as well as the challenges posed by reduced human control and weakened feedback loops:

\begin{itemize} [leftmargin=0.5cm]
    \item \textbf{Potential for End-to-End Solutions.}
    Generative AI search demonstrates significant potential for addressing complex, multi-faceted queries that require synthesis and creativity, surpassing the capabilities of traditional top-10 hyperlink-based results. However, this capability comes with the risk of generating inaccurate or hallucinated content, underscoring the need for robust mechanisms to ensure factual grounding and reliability.
    \item \textbf{Search Result Delivery.}
    Traditional search engines provide users with a list of links and snippets, enabling direct comparison and exploration of multiple sources. In contrast, generative AI search delivers a unified, synthesized answer, streamlining the user experience but potentially sacrificing transparency and human control for the entire search process.
    \item \textbf{User Feedback Mechanism.}
    Traditional search systems benefit from a rich and granular feedback loop, driven by user interactions such as clicks and dwell time. These signals enable continuous refinement of retrieval and ranking models. Generative AI search, however, typically relies on coarser feedback mechanisms, such as binary likes/dislikes or free-text corrections, which provide limited insights into specific failures within the pipeline (e.g., query decomposition, relevant document retrieval, or answer generation).
\end{itemize}

In summary, the evolution from traditional link-based retrieval systems to generative AI search, which provides direct answers, has undeniably enhanced user convenience and streamlined the search experience.
However, this shift introduces a \textbf{fundamental challenge}: the once-critical feedback loop that fueled constant improvements in Web search engines is now at risk of stalling. Generative search engines consolidate multiple user-driven interactions—such as decomposing information needs into sub-queries, selecting and examining multiple SERPs, and aggregating knowledge from retrieved documents—into an end-to-end pipeline. While this integration significantly improves usability, it simultaneously reduces the granularity of user feedback, making it difficult to diagnose and address specific weaknesses in the system. For instance, user dissatisfaction with the final answer could arise from various underlying issues, such as flawed task decomposition, inadequate retrieval, or errors in the LLM's summarization process. Yet,  feedback limited to the final output fails to provide the necessary insights to pinpoint the root cause of these failures across the complex pipeline. This limitation poses a significant barrier to iterative refinement and large-scale self-improvement, which have long been successful hallmarks of traditional search systems.

\section{NExT-Search Paradigm} \label{sec:NExT-Search_paradigm}
In this section, we present \textbf{NExT-Search}, a new paradigm aimed at rebuilding the user feedback ecosystem for generative AI search by reintegrating fine-grained user feedback throughout the whole search pipeline, akin to the success of traditional Web search. 

\subsection{Motivation and Overview}
As discussed in Section~\ref{sec:pipeline_comparison_analysis}, while generative AI search improves usability by offering end-to-end answers, it sacrifices the fine-grained feedback mechanisms that once enabled traditional Web search engines to improve iteratively. Most current systems only receive coarse signals, such as likes or dislikes for final answers, which provide little insight into which specific component—query decomposition, retrieval, or generation—led to user dissatisfaction.

To this end, \textbf{NExT-Search} is motivated by the need to reintroduce such fine-grained feedback without sacrificing the advanced user experience of modern generative AI search. As illustrated in Figure~\ref{fig:NExT-Search_pipeline}, NExT-Search integrates two complementary feedback mechanisms: \textbf{User Debug Mode} allows engaged users to intervene at various stages of the search pipeline—editing queries, re-ranking results, and refining generated answers—while \textbf{Shadow User Mode} employs a \emph{personalized user agent} to infer and simulate user preferences when explicit feedback is unavailable. By combining these modes, NExT-Search enables that every search interaction—whether fully guided by users or passively inferred—contributes structured signals for continuous optimization. 

In the following sections, we describe these two feedback mechanisms in detail (Section~\ref{sec:debug_mode} and Section~\ref{sec:shadow_mode}) with a concrete query example of \underline{``Plan a trip to attend SIGIR 2025''} and explore their synergy (Section~\ref{sec: synergy}) and introduce an incentive mechanism to encourage user participation (Section~\ref{sec: motivating}).

\subsection{User Debug Mode} \label{sec:debug_mode}
In \emph{User Debug Mode}, NExT-Search empowers users with deep, hands-on control over the generative AI search pipeline. By providing an interactive window and a transparent breakdown of each search stage, users can contribute fine-grained feedback that immediately affects system output (\emph{online adaption} in Section~\ref{sec:online_adaptation}) and is recorded for later model refinement (\emph{offline update} in Section~\ref{sec:offline_update}). This mode benefits scenarios where users have the motivation and expertise to debug and refine the search process, and it paves the way for more precise, interpretable, and adaptive generative AI search. In what follows, we detail how a user under this debug mode may intervene at each stage to debug the system.

\subsubsection{Debug in Query Decomposition} \label{sec: debug_qd}
Generative AI search engines initially attempt to break a user’s broad or ambiguous query into multiple sub-queries (or subtasks). 
However, several types of errors commonly arise at this stage:
\begin{itemize}[leftmargin=*]
    \item \emph{Misses key subtasks}: The system may overlook essential queries required to fulfill the user’s intent (e.g., omitting ``registration fees and process'' when planning to attend a conference).
    \item \emph{Includes irrelevant queries}: Extraneous queries may be introduced, leading to inefficiencies (e.g., including ``local sightseeing recommendations'' when the primary goal is attending a conference).
    \item \emph{Improperly orders subtasks}: Subtasks may be sequenced incorrectly, causing inefficiencies in information retrieval (e.g., searching for hotel bookings before confirming the schedule).
\end{itemize}

To tackle these issues and reintroduce fine-grained user signals, \emph{User Debug Mode} enables the following feedback mechanisms for user debugging in this stage:
\begin{itemize}[leftmargin=*]
    \item \emph{Add/Remove Sub-queries:} Users can introduce missing queries or remove redundant ones to align with their actual information needs (e.g., adding a query about ``registration fees'' or removing ``local sightseeing recommendations'').
    \item \emph{Reorder Sub-queries:} When task dependencies exist (e.g., the user must confirm the conference dates before booking flights), users can adjust the order of execution accordingly.
    \item \emph{Refine Constraints:} Users can fine-tune specific constraints within sub-queries, such as adjusting budget limits.
\end{itemize}

An example of debugging at this stage is shown below:

\begin{tcolorbox}[colback=gray!5!white,colframe=gray!75!black,title=User Debug in Query Decomposition]
\textbf{Initial Query Decomposition:}
\begin{itemize}[leftmargin=*]
    \item \textbf{[Q1]} What are the best flight options from [User's City] to the [conference location]?
    \item \textbf{[Q2]} Where and when will SIGIR 2025 be held?
    \item \textbf{[Q3]} What are the recommended hotels near the conference venue?
    \item \textbf{[Q4]} What are some sightseeing attractions near the conference venue?
\end{itemize}

\textbf{User Debugging:}
\begin{itemize}[leftmargin=*]
    \item \textcolor{black}{\textbf{Remove}} \textbf{[Q4]}: User's primary goal is to attend SIGIR 2025, and sightseeing is not a priority.
    \item \textcolor{black}{\textbf{Add}} \textbf{[Q5]}: ``What is the registration process and cost for SIGIR 2025?''— a prerequisite for travel planning.
    \item \textcolor{black}{\textbf{Reorder}} \textbf{[Q2] and [Q1]}: Adjust order to book travel only after confirming the event schedule.
\end{itemize}
\end{tcolorbox}

By logging each user edit (e.g., frequently added subtasks, repeatedly deleted constraints), the system amasses valuable data to refine its decomposition strategy over time. 


\subsubsection{Debug in Retrieval \& Ranking}
After decomposition, the system retrieves relevant documents or passages from indexed sources. However, common errors may occur:
\begin{itemize}[leftmargin=*]
    \item \emph{Retrieving irrelevant documents}: Results may include off-topic or outdated content.
    \item \emph{Missing high-quality or authoritative sources}: Important references such as official government websites or conference organizers’ pages might be underrepresented in the retrieved results.
    \item \emph{Suboptimal ranking}: The system may rank less useful documents higher than more relevant ones.
\end{itemize}

To address these retrieval and ranking issues while incorporating fine-grained user feedback, \emph{User Debug Mode} allows users to refine the retrieval and ranking process through several mechanisms:
\begin{itemize}[leftmargin=*]
    \item \emph{Annotate Relevance:} Users can explicitly mark documents as \emph{relevant}, \emph{partially relevant}, or \emph{irrelevant}.
    \item \emph{Re-rank Documents:} Users can manually adjust document priority, ensuring that the most useful sources are emphasized.
    \item \emph{Set Time/Domain Filters:} Users can refine retrieval criteria by restricting results to specific timeframes or limiting sources to specific domains (e.g., only include results from ``sigir.org'').
\end{itemize}

An example of debugging at this stage is shown below:

\begin{tcolorbox}[colback=gray!5!white,colframe=gray!75!black,title=User Debug in Retrieval \& Ranking]
\textbf{Initial Retrieved Results for sub-query: ``Where and when will SIGIR 2025 be held?''}
\begin{itemize}[leftmargin=*]
    \item \textbf{[D1]} News article on 2025 AI conferences briefly mentioning SIGIR (Partially relevant)
    \item \textbf{[D2]} SIGIR 2025 announcement from \textcolor{black}{ACM SIGIR website} (Highly relevant)
    \item \textbf{[D3]} 2023 SIGIR proceedings mentioning past conference locations (Irrelevant)
\end{itemize}

\textbf{User Debugging:}
\begin{itemize}[leftmargin=*]
    \item \textcolor{black}{\textbf{Remove}} \textbf{[D3]}: Outdated sources.
    \item \textcolor{black}{\textbf{Re-rank}} \textbf{[D2] above [D1]}: Prioritize official SIGIR sources over news articles.
    \item \textcolor{black}{\textbf{Apply domain filter}}: Restrict results to ``sigir.org'' to focus on authoritative sources.
\end{itemize}

\end{tcolorbox}

By logging each user edit (e.g., frequently excluded irrelevant documents, consistent source preferences, and re-ranking adjustments), the system accumulates valuable data to refine its retrieval and ranking models offline. 

\subsubsection{Debug in Answer Generation}
After retrieving and ranking relevant documents, the system synthesizes a response with LLMs. While this process provides a seamless, end-to-end search experience, it also introduces potential issues, including:
\begin{itemize}[leftmargin=*]
    \item \emph{Factual inaccuracies}: The model may generate hallucinated claims or misinterpret retrieved documents.
    \item \emph{Incomplete or excessive information}: The response might omit important details or contain unnecessary elaboration.
    \item \emph{Inappropriate style or tone}: The output may be too formal, too casual, overly technical, or lack proper structuring.
\end{itemize}

To empower users in refining the final response, \emph{User Debug Mode} introduces the following debugging mechanisms:
\begin{itemize}[leftmargin=*]
    \item \emph{Correct Factual Errors:} Users can highlight incorrect statements or ask for additional supporting evidence.
    \item \emph{Edit Partial Content:} Users can directly modify specific sections of the response by adding, removing, or restructuring content for improved clarity and accuracy.
    \item \emph{Adjust Style or Length:} Users can refine verbosity and tone based on their needs.
\end{itemize}

An example for debugging in this stage is shown below:

\begin{tcolorbox}[colback=gray!5!white,colframe=gray!75!black,title=User Debug in Answer Generation]
\textbf{Initial Generated Answer:} \\
``SIGIR 2025 will be held at the University of Padua, Italy, from July 15-19. The recommended conference hotels include NH Hotel Padova and Best Western Hotel Biri, with prices starting at 120\texteuro ~per night.''

\textbf{User Debugging:}
\begin{itemize}[leftmargin=*]
    \item \textcolor{black}{\textbf{Correct Factual Error:}} The conference dates and venue are incorrect; verify with the official SIGIR website.
    \item \textcolor{black}{\textbf{Edit Partial Content:}} Instead of listing only two hotels, request more budget-friendly options.
    \item \textcolor{black}{\textbf{Adjust Style or Length:}} Summarize the hotel details in bullet points for easy comparison.
\end{itemize}

\textbf{Revised Answer:} \\
``According to the official website, SIGIR 2025 will be hosted at the Padova Congress Center in Padua, Italy, from July 13–17. Recommended accommodations include:  
\begin{itemize}[leftmargin=*]
\item \emph{NH Hotel Padova} (120\texteuro/night, 10 min walk)   
\item \emph{Best Western Hotel Biri} (165\texteuro/night, 20 min walk)  
\item \emph{B\&B Hotel Padova} (90\texteuro/night, 10 min walk)  
\end{itemize}
For more lodging options, refer to the official SIGIR 2025 accommodation page: \url{https://sigir2025.dei.unipd.it/recommended-hotels.html}.''
\end{tcolorbox}

By logging frequent user interventions, such as recurring factual corrections, content modifications, or style adjustments, \emph{User Debug Mode} gathers valuable data to update for answer generation models, ensuring better alignment with user preferences over time.

\subsection{Shadow User Mode}
\label{sec:shadow_mode}

In scenarios where users prefer minimal interaction, \textbf{Shadow User Mode} employs a \emph{personalized user agent} to simulate user behavior and generate fine-grained feedback. This agent serves as an intelligent intermediary, assisting users who wish to refine their search process but find manual debugging cumbersome. By proactively suggesting AI-assisted feedback at each key stage, the agent reduces the interaction cost while still incorporating user preferences. As the agent continuously learns and improves its ability to mimic user decisions, it can increasingly replace direct user intervention, ensuring that generative AI search accumulates meaningful feedback without requiring extensive manual input.

The implementation of the \emph{personalized user agent} consists of two key components:
(1) \textbf{User Preference Learning}, which constructs and maintains a dynamic user profile based on past interactions to infer likely search preferences; 
(2) \textbf{AI-assisted Feedback Generation}, which predicts pseudo-feedback when users do not explicitly engage with intermediate steps.

\subsubsection{User Preference Learning} \label{sec: user_preference_learning}

The goal of user preference learning is to construct and continuously update a dynamic user profile that captures individual preferences and behavioral patterns. By analyzing various signals—including demographic attributes, search behaviors, click interactions, and browsing history—the system models user-specific tendencies to better align search outcomes with their expectations~\cite{xu2025personalized}. Conceptually, this process identifies overarching user preferences that influence search behavior. For example, if a user consistently prioritizes convenience over cost when booking accommodations, the system recognizes and encodes this preference. 
Ultimately, this phase will generate a structured user profile that guides the personalized user agent in providing more relevant and personalized AI-assisted feedback at each stage of the search pipeline, reducing unnecessary user effort while maintaining high-quality search refinements. 

\subsubsection{AI-assisted Feedback Generation} \label{sec: pseudo-feedback_generation}
With the constructed user profile, the \emph{personalized user agent} can assist users who prefer minimal interaction but still seek to refine their search results. When a user finds the generated answer unsatisfactory and wants to debug the process, the agent provides targeted correction suggestions based on their preferences, requiring only confirmation rather than manual intervention. This reduces user effort while ensuring valuable process-level feedback is continuously integrated. Next, we illustrate how the agent can generate intermediate feedback to assist users in debugging the search pipeline.

\textbf{Simulating User Feedback in Query Decomposition.}  
The personalized user agent leverages the learned user preferences to autonomously debug query decomposition by deciding whether to remove, reorder, or refine sub-queries. Instead of requiring the user to manually adjust them, the agent presents modification suggestions for confirmation. Once approved, it executes these refinements and provides a brief explanation of the changes. Below is a potential implementation prompt example:

\begin{tcolorbox}[colback=gray!5!white,colframe=blue!75!black,title=Simulating User Feedback in Query Decomposition]
\textbf{System Prompt:}  
You are simulating a user who wants to refine a query decomposition process. Based on the provided user profile, you will review the initial sub-queries and identify necessary adjustments. You can perform the following actions:
\{\textit{Description of the actions in this stage}\}

\textbf{User Profile:} \{\textit{User-specific preferences}\} \\
\textbf{User Query:} \{\textit{User query}\} \\
\textbf{Initial Query Decomposition:} \{\textit{Original sub-queries}\} \\
\textbf{Task Prompt:}  
Analyze the given sub-queries in light of the user profile, highlighting any necessary modifications with clear explanations. Then, generate a refined list of sub-queries that better align with the user's needs.
\end{tcolorbox}

\textbf{Simulating User Feedback in Retrieval \& Ranking.}  
In this stage, the personalized user agent refines the retrieval and ranking process based on learned user preferences and context. 
Instead of requiring users to manually sift through documents, the agent proactively suggests adjustments and presents a revised list for user confirmation. Below is a potential implementation prompt example:

\begin{tcolorbox}[colback=gray!5!white,colframe=blue!75!black,title=Simulating User Feedback in Retrieval \& Ranking]
\textbf{System Prompt:}  
You are simulating a user who wants to refine a retrieval \& ranking process. Based on the provided user profile, you will review the initial retrieved results and apply necessary adjustments. You can perform the following actions:  
\{\textit{Description of the actions in this stage}\}

\textbf{User Profile:} \{\textit{User-specific preferences}\} \\
\textbf{User Query:} \{\textit{User query}\} \\
\textbf{Initial Retrieved Results:} \{\textit{Original ranked document list}\} \\
\textbf{Task Prompt:}  
Analyze the retrieved documents in light of the user profile and context, identifying any necessary refinements with clear justifications. Then, generate a revised ranked list that best aligns with the user’s intent.
\end{tcolorbox}

\textbf{Simulating User Feedback in Answer Generation.}  
Finally, the agent refines the generated answer based on the user's learned preferences. Instead of requiring users to manually adjust factual correctness, content structure, or stylistic elements, the agent proactively suggests modifications aligned with their past preferences. 
Below is a potential implementation prompt example:

\begin{tcolorbox}[colback=gray!5!white,colframe=blue!75!black,title=Simulating User Feedback in Answer Generation]
\textbf{System Prompt:}  
You are simulating a user who wants to refine an AI-generated answer. Based on the provided user profile, you will review the initial answer and apply necessary adjustments. You can perform the following actions:  
\{\textit{Description of the actions in this stage}\}

\textbf{User Profile:} \{\textit{User-specific preferences}\} \\
\textbf{User Query:} \{\textit{User query}\} \\
\textbf{Initial Generated Answer:} \{\textit{Original generated answer}\} \\
\textbf{Task Prompt:}  
Analyze the generated answer in light of the user profile and query context, highlighting necessary modifications with clear justifications. Then, generate a revised answer that best aligns with the user’s needs.
\end{tcolorbox}

\subsection{Synergy of Dual Feedback Modes} \label{sec: synergy}
Both \emph{User Debug Mode} and \emph{Shadow User Mode} share the common objective of maintaining a continuous flow of user-driven signals to support both \emph{online} (current session) and \emph{offline} (long-term) model refinement. In \emph{User Debug Mode}, engaged users can exert full control over the search pipeline by modifying sub-queries, re-ranking retrieved documents, and refining generated answers. However, this increased level of human control comes at the cost of greater interaction complexity. \emph{Shadow User Mode} mitigates this by deploying a personalized user agent that learns user behaviors and provides fine-grained AI-assisted feedback when explicit debugging is absent. As the agent progressively refines its ability to simulate user preferences, it can deliver increasingly high-quality feedback, reducing the need for manual intervention. Over time, users can gradually delegate more of the debugging process to the agent, trusting it to make refinements on their behalf.  

This synergy ensures that every search session—whether actively debugged or passively simulated—contributes valuable feedback for improving the generative AI search system. \emph{User Debug Mode} provides high-fidelity ``gold'' signals when users directly interact with the pipeline, while \emph{Shadow User Mode} supplies continuous AI-assisted feedback in cases of minimal engagement. Together, these modes reconstruct a robust, multi-stage feedback ecosystem: immediate user corrections address urgent errors, while aggregated logs—comprising both explicit user input and agent-generated AI-assisted feedback—fuel iterative improvements to query decomposition, ranking, and generation modules.

\subsection{Motivating User Engagement} \label{sec: motivating}

Encouraging users to actively participate in NExT-Search's feedback mechanisms is crucial for maintaining a robust and continuously improving generative AI search ecosystem. However, engaging users in deep debugging processes requires additional effort. Without clear incentives, users may be reluctant to invest the necessary time and cognitive resources. To address this, we introduce a \textbf{feedback store} as a promising mechanism to motivate user participation. In this marketplace, users can package their optimized debugging processes into reusable templates and offer them to others facing similar search challenges. These templates can be listed for purchase, where contributors receive direct financial compensation when others adopt their solutions, or they can generate passive income based on usage metrics such as views, downloads, or successful query resolutions.

This feedback store creates a closed-loop knowledge monetization loop, enabling experienced users to capitalize on their expertise while allowing less experienced users to benefit from high-quality, pre-optimized search workflows without the need for manual refinement. By bridging the gap between expert contributors and general users, the feedback store fosters a self-sustaining ecosystem for search refinement and continuous improvement. Users not only achieve more efficient and accurate search outcomes but also receive tangible incentives, ultimately creating a mutually beneficial dynamic between the platform and its contributors.

\section{Leveraging Feedback: From Online to Offline} \label{sec: feedback_use}

\begin{figure}[t]  
    \centering    
    \includegraphics[width=1\linewidth]{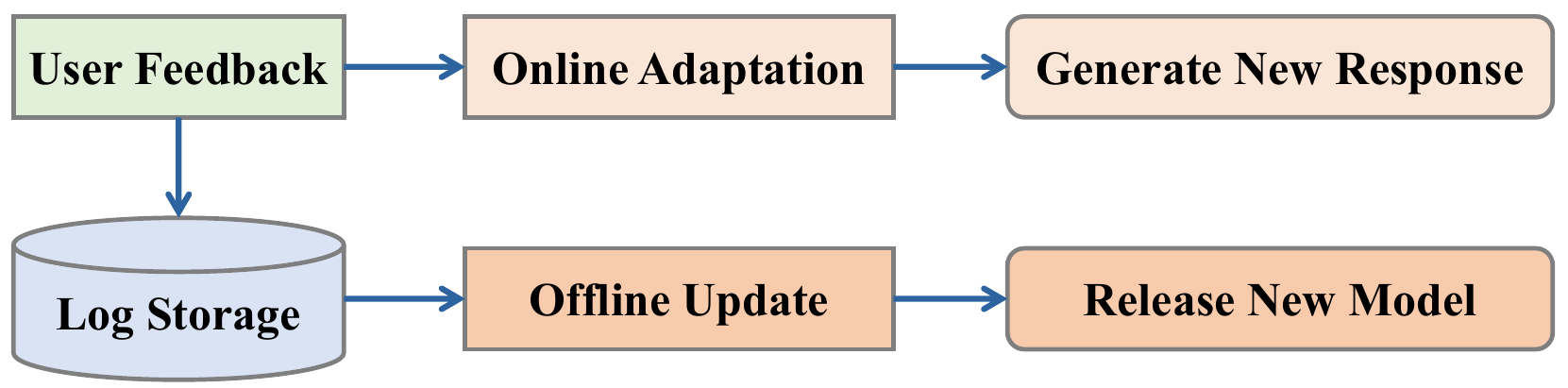}
    \caption{Two mechanisms for leveraging feedback.}
    \label{fig:update}  
    \vspace{-0.1cm}
\end{figure}

With the step-by-step feedback collected under our NExT-Search paradigm, we explore two complementary strategies to leverage the feedback in Figure~\ref{fig:update}: \textbf{Online Adaptation}, which refines responses \emph{in real time} based on user feedback within the current active session, and \textbf{Offline Update}, where accumulated interaction logs drive model retraining, reinforcing \emph{a long-term self-improvement loop}.

\subsection{Online Adaptation} \label{sec:online_adaptation}
Online adaptation focuses on dynamically improving response quality during an ongoing user session. As feedback is provided—whether through explicit corrections in \emph{User Debug Mode} or inferred AI-assisted feedback in \emph{Shadow User Mode}—the system applies immediate refinements to better align with the user's intent. Once a user modifies any stage in the search pipeline, all subsequent stages are re-executed accordingly, akin to debugging a program where each adjustment propagates downstream to ensure consistency.

For example, when users modify sub-queries—such as adding a missing query—the system immediately reprocesses the updated formulation, ensuring that all downstream stages reflect the changes. When users annotate retrieved documents for relevance or apply filtering criteria, the system dynamically re-ranks results, improving the quality of the knowledge pool before answer synthesis. Finally, if users correct factual errors or request additional details in the generated response, the system selectively regenerates affected sections while preserving validated content, reducing unnecessary recomputation and enhancing efficiency. Through these real-time adaptations, NExT-Search allows generative AI search to continuously align with evolving user intent.

\subsection{Offline Update} \label{sec:offline_update}

Beyond immediate corrections, offline update aggregates multi-session interaction logs to drive long-term system improvements. User feedback—both explicit and simulated—serves as structured supervision signals for continuously refining key components of the generative AI search pipeline. In industrial search systems, a widely adopted strategy is \emph{daily incremental updates}~\cite{liu2009learning, manning2009introduction, levene2011introduction}, where user interaction and feedback logs are periodically processed to generate positive and negative training samples. These samples are then used to perform incremental training on top of the previous day’s model parameters. Once training is completed, the updated model is deployed to production.

For each of the three core stages of the pipeline, we discuss how offline update can be effectively constructed using user feedback:

\textbf{Update for Query Decomposition.}
User-corrected sub-queries provide direct supervision signals for improving query decomposition. The system collects pairs of \emph{original sub-queries} (before correction) and \emph{revised sub-queries} (after user modification), treating the latter as positive examples and the former as negative examples. These structured samples can then be used to refine the LLM’s decomposition abilities through techniques such as instruction fine-tuning~\cite{ouyang2022training} or direct preference optimization~\cite{rafailov2024direct}.

\textbf{Update for Retrieval \& Ranking.}
Feedback from user annotations—including relevance labels, source preferences, and re-ranking actions—serves as essential signals for improving retrieval and ranking models. Positive samples consist of documents that users frequently engage with (e.g., those marked as relevant, clicked, or cited in responses), while negative samples include documents that users downvoted or explicitly filtered out. These signals are then used to fine-tune retrieval models~\cite{zhao2022dense} and ranking models~\cite{xu2018deep} to better reflect user preferences.

\textbf{Update for Answer Generation.}
Corrections made to generated responses, such as factual fixes or content expansions, are logged as supervised learning signals for improving the LLM’s answer synthesis capabilities. Positive samples include sections that users accepted or minimally modified, while negative samples are those that were corrected or flagged as hallucinations. 
These signals can be used to fine-tune the LLM—e.g., through reinforcement learning from human feedback (RLHF)~\cite{ouyang2022training, bai2022training}—to improve factual accuracy and better align responses with user expectations.

By continuously leveraging these structured feedback signals, offline update enables generative AI search to iteratively refine the search pipeline over the long term.

\section{Potential Research Opportunities} \label{sec:future_opportunities}

Under \textbf{NExT-Search} paradigm, several open challenges and research directions emerge, offering rich avenues for future investigation. 
We outline the following research opportunities:

\textbf{LLM for Personalized User Simulation.}
One critical challenge in realizing the full potential of \emph{Shadow User Mode} under the NExT-Search paradigm lies in building personalized user agents capable of reliably simulating user behavior and producing high-quality, fine-grained feedback from limited user interaction data. Future work may explore advanced user modeling techniques that combine large reasoning models with behavioral data to infer preferences more accurately~\cite{xu2025personalized}. 
Promising directions include combining RAG with personalized in-context learning~\cite{salemi2024comparing}, constructing dynamic memory modules~\cite{zhang2024personalized} to retain user-specific history across sessions, and leveraging preference-conditioned generation to align feedback suggestions with individual goals.
Furthermore, building such simulators often requires access to sensitive signals such as user profiles, engagement logs, or contextual attributes, which may raise privacy concerns~\cite{shen2007privacy, dai2024bias, Balog2025UserSI}. As a result, balancing personalization with privacy is another fundamental challenge. Techniques such as federated learning~\cite{Tan2021TowardsPF} and on-device adaptation~\cite{mendoza2024adaptive} could offer promising pathways for privacy-preserving user simulation.

\textbf{Learning from Human and AI-Assistant Feedback.}
While the NExT-Search paradigm envisions a renewed user feedback ecosystem for generative AI search, effectively leveraging the collected signals to drive system improvement remains an open challenge. In Section~\ref{sec: feedback_use}, we outline two complementary strategies: \emph{online adaptation} for immediate refinements and \emph{offline updates} for longer-term model retraining. However, how to efficiently implement these strategies—particularly in leveraging fine-grained user feedback—deserves deeper exploration in future work. One promising direction is to design training procedures that utilize users' step-by-step feedback trajectories to supervise various components of the search pipeline. Integrating recent advances in LLM reasoning-aware training~\cite{Li2025FromS1} may help construct richer learning strategies and improve pipeline robustness. Moreover, another core challenge lies in integrating heterogeneous feedback sources. While \emph{User Debug Mode} offers high-quality but sparse feedback and \emph{Shadow User Mode} supplies abundant but potentially noisy signals, effectively combining these complementary signals remains a key research challenge. Thus, exploring adaptive learning techniques such as multi-task learning~\cite{ma2018modeling} or curriculum learning~\cite{bengio2009curriculum} may offer promising avenues for training on different feedback types. More broadly, the question of how to maximize system robustness and learning efficiency from diverse feedback streams is central to realizing the full potential of the NExT-Search paradigm.

\textbf{Human-Centric Interaction Design.}
Although \emph{Shadow User Mode} is an effective complement to \emph{User Debug Mode}—using LLMs to proactively suggest feedback and reduce user effort—the paradigm still fundamentally depends on user engagement. Thus, the system must strike a careful balance between transparency, control, and interaction burden. 
This raises several important design questions: How should intermediate steps (e.g., sub-query decomposition or retrieved documents) be presented to encourage actionable feedback? What level of user intervention is appropriate across different users and tasks?
Collaboration with researchers from human-computer interaction and user behavior studies~\cite{Bates1989TheDO, Jiang2014SearchingBA, Spatharioti2023ComparingTA} could yield innovative UI designs or interactive workflows that solicit targeted feedback, minimize user frustration, and gradually train novices to handle more complex tasks.
Another open challenge is how to dynamically manage transitions between \emph{Shadow User Mode} and \emph{User Debug Mode}. Developing adaptive mode-switching mechanisms—based on task complexity, user expertise, or predicted benefit-to-cost ratios—presents a promising research direction. Techniques such as user modeling~\cite{shen2005implicit} or reinforcement learning~\cite{yao2020rlper} could be leveraged to personalize interaction strategies.

While our NExT-Search paradigm offers an initial step toward rethinking feedback in generative AI search, future work should investigate how feedback mechanisms can be continuously refined and adapted in response to the rapidly evolving pipelines and interaction patterns of generative search systems.


\section{Conclusion}
In this work, we envision \textbf{NExT-Search}, a paradigm aimed at reintroducing fine-grained user feedback into generative AI search. By integrating \emph{User Debug Mode} for direct user interventions and \emph{Shadow User Mode} for implicit feedback simulation, NExT-Search enables the collection of structured, stage-level signals across the entire search pipeline. These feedback signals can be further leveraged through online adaptation for real-time answer refinements and offline update for long-term model improvements. Additionally, we introduced a feedback store mechanism to motivate user engagement. 
While NExT-Search remains a forward-looking framework, we believe it offers valuable insights for the design of next-generation generative AI search, ensuring a balance between automation, user control, and continuous self-improvement. Due to the lack of publicly available datasets, we leave empirical validation and system implementation to future work.

\begin{acks}
This research/project is supported by the National Research Foundation, Singapore under its National Large Language Models Funding Initiative, (AISG Award No: AISG-NMLP-2024-002), the National Natural Science Foundation of China (No.62276248), and the Youth Innovation Promotion Association CAS under Grants (No.2023111). 
This research is supported by A*STAR, CISCO Systems (USA) Pte. Ltd and National University of Singapore under its Cisco-NUS Accelerated Digital Economy Corporate Laboratory (Award I21001E0002).
Any opinions, findings, conclusions, or recommendations expressed in this material are those of the author(s) and do not reflect the views of National Research Foundation, Singapore.
\end{acks}

\balance
\bibliographystyle{ACM-Reference-Format}
\bibliography{ref}

\end{document}